\documentclass[12pt,a4paper]{article}
\usepackage{latexsym}

\begin{document}
 
\title{Renormalization Group Flow in BRST Invariant Open String $\sigma$-Model}
\author{Rui Neves\thanks{Research supported by
F.C.T.'s PRAXIS XXI
Post-Doctoral Fellowship BPD/14137/97.}\\
\footnotesize{\'Area Departamental de F\'{\i}sica, UCE,
Universidade do Algarve}\\
\footnotesize{Campus de Gambelas, 8000 Faro, Portugal}\\
\footnotesize{rneves{\it @}mozart.ualg.pt}}
\maketitle

\begin{abstract}
The renormalization group flow in the theory space of a BRST invariant
string $\sigma$-model is investigated. For the open bosonic string the
non-perturbative off-shell effective action and its gauge symmetry
properties are determined from $\beta$-functions defined by the local Weyl
anomaly. The interactions are shown to explicitly break the free
theory BRST invariance generating new non-linear gauge symmetries of
the type present in Witten's string field theory. In the
Feynman-Siegel gauge the $\sigma$-model is shown to generate Witten's 
structure of vertex couplings. 
\end{abstract}

\section{Introduction}

Conformal invariance on the world-sheet is at the heart of string
theory. One of its fundamental consequences is that the
dynamics of strings is constrained to curved spacetime backgrounds
which are solutions of the string equations of motion. These conformal
symmetry conditions are equivalent to the vanishing of the non-linear
$\sigma$-model $\beta$-functions and so classical string vacua as well
as on-shell string scattering amplitudes are characteristics of the
renormalization group (RG) fixed points. This was first discovered when
studying string dynamics on massless background fields \cite{SMML} and
rapidly seen to hold when including condensates of other string modes
\cite{DS,AO,ZP,BM,BS,BNS,KS,AR,HLP,EFE,LV,JJ}.

The conformal fixed points belong to a RG flow in an infinite dimensional 
field theory space spanned by all the string $\sigma$-model couplings.
As indicated by the investigations in the vicinity of the on-shell
gaussian fixed point, the RG flow looks like a gradient flow generated by 
a background field effective action $I$ and a theory space metric 
$G^{jl}$ (see for example \cite{ZP,BM,KS,HLP}). The
$\beta$-function $\beta^j$ corresponding to the coupling $g^j$ is then
given by 

\vspace{-0.5cm}
\begin{equation}
{\beta^j}={G^{jl}}{{\delta I}\over{\delta{g^l}}}\,.
\end{equation}
Since the effective action $I$ is to be considered an off-shell
functional of all string couplings it may naturally be interpreted as 
the tree level action for a string field theory.  

The precise non-perturbative definition of the possible spacetime effective
actions $I$ associated with the RG flow is still an open problem. At the
free field theory level the early uncovering of the $L_{-1}$ string
gauge invariances has not yet been followed by an equally clear successful
generation of the $L_{-n}$, $n\geq 2$ gauge symmetries
\cite{BS,AR,HLP,LV,JJ,BFLP,BB}. At the level of the
interactions a near mass-shell field redefinition ambiguity was found 
unavoidable when deducing the covariant form of gauge fixed actions \cite{AT}. 
Without the field redefinitions the non-perturbative interactions were 
shown to depend on the chosen regulator \cite{SE,KPP} and a
scheme \cite{KPP} was found for which the structure of Witten's string field
theory (WSFT) and the associated non-linear gauge symmetry \cite{EW,KoS} are 
obtained from the $\sigma$-model RG flow.

In this letter we consider the
introduction of ghost field couplings in the $\sigma$-model based on the
approach of Jain and Jevicki \cite{JJ}. To study we
select the open bosonic string and the region of theory space
corresponding to the tachyon $T$, the photon $A^\mu$ and the ghost
field $\alpha$. We construct a BRST invariant string $\sigma$-model
and determine the 
non-perturbative off-shell effective action $I$ and its 
gauge symmetry properties. To be able to do so we define the $\beta$-functions by the local Weyl anomaly calculated
via the covariant heat kernel regulator. We also require a unique but 
non-local theory space metric $G^{jl}$. We show that the interactions
explicitly break the free field theory BRST invariance generating
new non-linear gauge symmetries of the type present in WSFT. We also show that in the Feynman-Siegel gauge
$\alpha=0$ the effective action $I$ has Witten's
structure of vertex couplings. Decoupling the ghost sector the theory
space metric becomes trivial and the curved world-sheet $\sigma$-model 
for $T$ and $A^\mu$ still generates the structure
of Witten's theory without the need for the unsatisfactory
regularization scheme recently used in the literature \cite{KPP}.    

\section{The BRST Invariant String $\sigma$-Model}                 

As was proposed by Jain and Jevicki \cite{JJ} the spacetime ghost background fields
may be introduced in the string $\sigma$-model if the
reparametrization ghosts are bosonized and then coupled to the matter
sector. Considering the open bosonic string in $d$ dimensions and 
the theory space
associated with $T$, $A^\mu$ and $\alpha$, the 2D field theory
action $S$ is defined as $S={S_0}+{S_{\mbox{\scriptsize B}}}$ where

\vspace{-0.5cm}  
\begin{equation}
{S_0}={1\over{4\pi\alpha'}}\int{d^2}\tilde{s}\left({X^\mu}\tilde{\Delta}{X_{\mu}}+Y\tilde{\Delta}Y\right)+{Q\over{4\pi\sqrt{\alpha'}}}\left(\int{d^2}\tilde{s}\tilde{R}Y+2\int d\tilde{s}{k_{\tilde{g}}}Y\right),\label{1}        
\end{equation}
and

\vspace{-0.75cm}
\begin{equation}
{S_{\mbox{\scriptsize B}}}={g\over{2\sqrt{2\alpha'}}}
\int
d\tilde{s}\left\{\sqrt{2}\,T(X)+i{\partial_{\tilde{t}}}{X^\mu}{A_\mu}(X)+{i\over{Q}}
{\partial_{\tilde{t}}}Y\left[\sqrt{\alpha'}\,\partial\cdot A(X)-\alpha(X)
\right]\right\}.\label{2}
\end{equation}
As is clear this action has been constructed to be invariant
under the following BRST transformation of the bare field couplings 
($\Box={\partial^\mu}{\partial_\mu}$)

\vspace{-1.0cm}
\begin{eqnarray}
\delta T(X)&=&0,\nonumber\\
\delta{A^\mu}(X)&=&\sqrt{\alpha'}\,{\partial^\mu}\Lambda(X),\nonumber\\
\delta\alpha(X)&=&\alpha'\Box \Lambda(X).\label{3}
\end{eqnarray}

\vspace{-0.5cm}
\noindent This classical property is clearly a consequence of the $U(1)$
breaking background field interaction 
which couples the matter and ghost sectors. It is independent of the
normalization constants chosen in the boundary action 
$S_{\mbox{\scriptsize B}}$ which are justified by the quantum theory.  

In formulae (\ref{1}), (\ref{2}) and (\ref{3}) $\alpha'$ is the
standard Regge slope and $g$ the string coupling. On the world-sheet $X^\mu$ and $Y$ are taken to satisfy Neumann
boundary conditions,
${\partial_{\tilde{n}}}{X^\mu}=0={\partial_{\tilde{n}}}Y$. The
world-sheet is a 2D surface with $\tilde{R}$ and $k_{\tilde{g}}$ as its scalar and geodesic curvatures. Its metric is in the conformal
gauge,
${\tilde{g}_{ab}}(\xi)=\exp\left[2\varphi(\xi)\right]{\delta_{ab}}$,
where the coordinates $\xi=(\tau,\zeta)$, $-\infty<\tau<+\infty$, 
$0\leq\zeta<+\infty$ span the upper-half plane and $\varphi$ is the
Liouville field also satisfying Neumann boundary conditions ${\partial_{\tilde{n}}}\varphi=\exp\left[-\varphi(\tau)\right]{\partial_\zeta}\varphi(\xi){|_{\mbox{\scriptsize $\zeta=0$}}}=0$. The
integration elements are ${d^2}\tilde{s}={d^2}\xi\sqrt{\tilde{g}}={d^2}\xi\exp\left[2\varphi(\xi)\right]$
and $d\tilde{s}=d\tau\exp\left[\varphi(\tau)\right]$. The kinetic
operator is the covariant laplacian
$\tilde{\Delta}=-(1/\sqrt{\tilde{g}}){\partial_a}\sqrt{\tilde{g}}{\tilde{g}^{ab}}{\partial_b}=-\exp\left[-2\varphi(\xi)\right]{\partial^2}$
  and
  ${\partial_{\tilde{t}}}=\exp\left[-\varphi(\tau)\right]{\partial_\tau}$ is the derivative tangent to the boundary. The 2D surface curvatures are linked
  by the Gauss-Bonnet theorem $\int{d^2}\tilde{s}\tilde{R}+2\int
  d\tilde{s}{k_{\tilde{g}}}=4\pi\chi$, where for the disc topology
  $\chi=1$. With the topological background charge at the origin in
  the boundary $\tilde{R}=-2\exp\left[-2\varphi(\xi)\right]{\partial^2}\varphi(\xi)$
  and
  ${k_{\tilde{g}}}=2\pi\chi\exp\left[-\varphi(\tau)\right]\delta(\tau)$.  

The curvature coupling $Q$ is determined by the local Weyl anomaly
associated with the reparametrization ghosts of Polyakov's open 
bosonic string \cite{PM,MN}. Integrating $Y$ and using the covariant
heat kernel regulator we obtain 

\[
{\delta_\rho}\ln\left[{{\left({{\mbox{Det}'\tilde{\Delta}}\over{\int{d^2}\tilde{s}}}\right)}^{-1/2}}{e^{\tilde{\mathcal{F}}_Q}}\right]={{1+3{Q^2}}\over{24\pi}}\left(\int{d^2}\tilde{s}\tilde{R}\rho+2\oint
  d\tilde{s}{k_{\tilde{g}}}\rho\right)-
\]
\begin{equation}
-{Q^2}\chi{{\int{d^2}\tilde{s}\rho}\over{\int{d^2}\tilde{s}}}-{{2Q\chi}\over{\sqrt{\alpha'}\int{d^2}\tilde{s}}}\int{d^2}\tilde{s}(\xi){d^2}\tilde{s}(\xi')\rho(\xi){\tilde{G}_N}(\xi,\xi'){\tilde{J}_Q}(\xi'),\label{4}
\end{equation}
where the contributions associated with local
renormalization counterterms have been omited as they are ultimately
tuned to zero. In the Weyl anomaly (\ref{4}) we wrote

\vspace{-0.5cm}
\begin{equation}
{\tilde{\mathcal{F}}_Q}={1\over{4\pi\alpha'}}\int{d^2}\tilde{s}(\xi){d^2}\tilde{s}(\xi'){\tilde{J}_Q}(\xi){\tilde{G}_N}(\xi,\xi'){\tilde{J}_Q}(\xi'),
\end{equation}
where the functional current $\tilde{J}_Q$ verifies   

\vspace{-0.25cm}
\begin{equation}
{\tilde{J}_Q}={{Q\sqrt{\alpha'}}\over{2}}\left(\tilde{R}+
2{{\tilde{\delta}}_B^2}{k_{\tilde{g}}}\right),\quad\int{d^2}\tilde{s}{\tilde{J}_Q}=2\pi
Q\sqrt{\alpha'}\,\chi
\end{equation} 
and $\tilde{G}_N$ is the Neumann laplacian Green's function defined to
satisfy

\vspace{-0.25cm}
\begin{equation}
\tilde{\Delta}{\tilde{G}_N}(\xi,\xi')={{{\delta^2}(\xi-\xi')}\over{\sqrt{\tilde{g}}}}-{1\over{\int{d^2}\tilde{s}}},\quad{\partial_{\tilde{n}}}{\tilde{G}_N}(\xi,\xi')=0
\end{equation}
as well as to be symmetric in its arguments and orthogonal to the
laplacian's constant zero mode

\vspace{-0.25cm}
\begin{equation}
\int{d^2}\tilde{s}(\xi){\tilde{G}_N}(\xi,\xi')=0.
\end{equation}
The non-local piece in anomaly (\ref{4}) is generated by $\tilde{G}_N$
due to the presence of the laplacian's zero mode \cite{MN}. This term is zero
for a flutuating world-sheet with infinite area. Only then the anomaly 
(\ref{4}) may be equal to the Weyl anomaly associated
with the integration of the reparametrization ghosts in Polyakov's
open bosonic string. This is the condition for bosonization in this theory 
and it further selects $Q=\pm 3i$. Taking the infinite 
target space to have the critical $26$ dimensions the local
Weyl anomaly is also cancelled.

\section{The Renormalization Group Flow $\beta$-Functions}

The quantum features of the BRST invariant string $\sigma$-model we are
going to analyze in this work are defined by the RG flow
$\beta$-functions. Let us introduce them following the approach of
Klebanov and Susskind \cite{KS}. A first point to be noted is that the 
Liouville mode $\varphi$ takes the role played by the scale parameter
$t$. This
is because the global scale transformations are now extended to the
the Weyl transformations. Then the general Wilson 
RG equations for the set of renormalized fields $g^j$ are the 
following variational equations in $\varphi$

\vspace{-0.5cm}
\begin{equation}
{\beta^j}={{\delta{g^j}}\over{\delta\varphi}}={\lambda_j}{g^j}+
{\alpha^j_{kl}}{g^k}{g^l}+\cdots.\label{5}
\end{equation}
In Eqs. (\ref{5}) $\lambda_j$ are anomalous dimensions and 
$\alpha^j_{kl}$ vertex coefficients. The ellipsis represent higher
order vertex terms in the weak field expansion (WFE) of the
$\beta$-functions. Since the anomalous
dimension matrix is supposed to be diagonalized there is no summation
in $j$. The solutions of Eqs. (\ref{5}) in terms of the bare fields 
${g^j}(0)$ are 

\vspace{-0.75cm}
\begin{equation}
{g^j}(\varphi)=\exp\left({\lambda_j}\varphi\right){g^j}(0)+\left\{\exp\left[({\lambda_k}+{\lambda_l})\varphi\right]-\exp\left({\lambda_j}\varphi\right)
\right\}{{\alpha^j_{kl}}\over{{\lambda_k}+{\lambda_l}-{\lambda_j}}}{g^k}(0){g^l}(0)+\cdots.\label{6}
\end{equation}
       
So, to find the $\beta$-functions we have to determine the renormalized fields
($\hat{T}$, $\hat{A}^\mu$, $\hat{\alpha}$) as
functions of the bare fields ($T$, $A^\mu$, $\alpha$) and then compare
with Eqs. (\ref{5}) and (\ref{6}). We start by separating $X^\mu$ into
a classical
background $X_0^\mu$ and a quantum flutuation $\bar{X}^\mu$,
${X^\mu}={X_0^\mu}+{\bar{X}^\mu}$, where naturally both fields obey
Neumann boundary conditions. Taking the ghost coordinate $Y$ as a pure quantum
field we then consider in the WFE the $\sigma$-model partition function   

\vspace{-0.75cm}
\begin{equation}
Z=<1-{S_{\mbox{\scriptsize B}}}+{1\over{2}}{S_{\mbox{\scriptsize B}}^2}+\cdots>=
\int{\mathcal{D}_{\tilde{g}}}(\bar{X},Y)\exp
\left(-{S_0}\right)\left(1-{S_B}+
{1\over{2}}{S_B^2}+\cdots\right)
\end{equation}
and the 2D effective action ${S_{\mbox{\scriptsize eff}}}=-\ln Z$.

Now, Fourier transforming the bare background fields to momentum
space generates a sum of standard gaussian integrals. When integrated
these will produce extra dependence on
the conformal mode $\varphi$ due to the quantum Weyl anomalies associated with the functional contractions. Let us first consider the linear order of the WFE,

\vspace{-0.75cm}
\begin{equation}
<{S_{\mbox{\scriptsize B}}}>={g\over{2\sqrt{2\alpha'}}}{\int_{-\infty}^{+\infty}}
d\tau\int{d^{26}}k\int{\mathcal{D}_{\tilde{g}}}(\bar{X},Y)\exp\left\{-{S_0}+ik\cdot\left[{X_0}(\tau)+\bar{X}(\tau)\right]\right\}{L_{\mbox{\scriptsize B}}}(k,\tau),
\end{equation}
where

\vspace{-0.75cm}
\begin{equation}
{L_{\mbox{\scriptsize B}}}(k,\tau)=\sqrt{2}\exp\left[\varphi(\tau)\right]T(k)+i\left[{\dot{X}_0^\mu}(\tau)+
{\dot{\bar{X}}^\mu}(\tau)\right]{A_\mu}(k)+{i\over{Q}}\dot{Y}(\tau)\left[i\sqrt{\alpha'}\,k\cdot A(k)-\alpha(k)\right]
\end{equation}
and the dot over the fields is the notation for
${\partial_\tau}$. Here the extra dependence on $\varphi$ comes from the heat kernel regulated Neumann Green's function $\tilde{G}_N^R$ at coincident points \cite{MN},    

\vspace{-0.5cm}
\begin{equation}
{\delta_\rho}{\tilde{G}_N^R}(\tau,\tau)={1\over{\pi}}\rho(\tau),\label{7}
\end{equation}
and also from its first derivative

\vspace{-0.5cm}
\begin{equation}
{\delta_\rho}{\partial_\gamma}{\tilde{G}_N^R}(\gamma,\tau)
{\Big |_{\mbox{\scriptsize $\gamma=\tau$}}}={1\over{2\pi}}\dot{\rho}(\tau).\label{8}
\end{equation}

In the next order  we need to evaluate the contractions, 

\vspace{-0.1cm}
\[
<{1\over{2}}{S_{\mbox{\scriptsize B}}^2}>={{g^2}\over{16\alpha'}}
{\int_{-\infty}^{+\infty}}d{\tau_1}{\int_{-\infty}^{+\infty}}d{\tau_2}\int
{d^{26}}{k_1}\int{d^{26}}{k_2}\int{\mathcal{D}_{\tilde{g}}}(\bar{X},Y)
\exp(-{S_0})\times
\]
\begin{equation}
\times\exp
\left\{i{k_1}\cdot\left[{X_0}({\tau_1})+\bar{X}({\tau_1})\right]+
i{k_2}\cdot\left[{X_0}({\tau_2})+\bar{X}({\tau_2})\right]\right\}{L_{\mbox{\scriptsize B}}}({k_1},{\tau_1}){L_{\mbox{\scriptsize B}}}({k_2},{\tau_2}),
\end{equation}
which are associated with the
first two terms in the expansions of $[\varphi({\tau_2})$, 
${X_0^\mu}({\tau_2})]$ around $\tau_1$, 

\vspace{-0.5cm}
\begin{equation}
\left[\varphi({\tau_2}),{X_0^\mu}({\tau_2})\right]=\left[
\varphi({\tau_1}),{X_0^\mu}({\tau_1})\right]+\left[\dot{\varphi}({\tau_1}),
{\dot{X}_0^\mu}({\tau_1})\right]({\tau_2}-{\tau_1})+\cdots.
\end{equation}
All the terms with higher order derivatives of $\varphi$ or $X_0^\mu$
will be ignored as they only contribute to the renormalizations
of higher level massive fields.

In all the relevant contractions there are Weyl
anomalous divergences which introduce extra dependence on
$\varphi$. Besides those like (\ref{7}) and (\ref{8}) we also find
divergent integrals of the form 

\vspace{-0.75cm}
\begin{equation}
{\int_{-\infty}^{+\infty}}d{\tau_1}{\int_{-\infty}^{\tau_1}}d{\tau_2}
\exp\left[-\pi P({k_1},{k_2}){\tilde{G}_N}({\tau_1},{\tau_2})\right],\label{10}
\end{equation}
where
${\tilde{G}_N}({\tau_1},{\tau_2})={G_N}({\tau_1},{\tau_2})=-(1/\pi)\ln(|{\tau_1}-{\tau_2}|/\sqrt{\alpha'}\,)$.
Noting that for non-coincident points the Neumann Green's function does
not have a local Weyl anomaly \cite{MN} and using the
covariant heat kernel result (\ref{7}) we obtain the
following extra dependence on the Liouville mode $\varphi$

\vspace{-0.5cm}
\begin{equation}
-
\sqrt{\alpha'}{\int_{-\infty}^{+\infty}}d{\tau_1}{{\left[P({k_1},{k_2})+
1\right]}^{-1}}\exp\left\{-\left[P({k_1},{k_2})+1\right]\varphi({\tau_1})\right\}.
\end{equation}
Note as well that the integrals (\ref{10}) have a potential infrared
divergence appearing as ${\tau_2}\to-\infty$. To remove it we impose the 
convergence condition $P({k_1},{k_2})+1<0$ \cite{KS}. 

Then we find   

\vspace{-0.75cm}
\begin{equation}
{S_{\mbox{\scriptsize eff}}}={1\over{4\pi\alpha'}}
\int{d^2}\xi{\partial_a}{X_0^\mu}{\partial^a}{X_{0\mu}}+
{g\over{2\sqrt{2\alpha'}}}{\int_{-\infty}^{+\infty}}
d\tau\int{d^{26}}k\exp\left[ik\cdot{X_0}(\tau)\right]{\hat{\mathcal{L}}_{\mbox{\scriptsize B}}}(k,\tau)+\cdots,
\end{equation}
with 

\vspace{-0.75cm}
\begin{equation}
{\hat{\mathcal{L}}_{\mbox{\scriptsize B}}}(k,\tau)=\sqrt{2}\,\hat{T}\left[k,\varphi(\tau)\right]+
i{\dot{X}_0^\mu}{\hat{\mathcal{A}}_\mu}\left[k,\varphi(\tau)\right],
\end{equation}
where the renormalized fields are

\vspace{-0.75cm}
\begin{eqnarray}
\hat{T}\left[k,\varphi(\tau)\right]&=&\exp\left[(1-\alpha'{k^2})\varphi(\tau)\right]\bigg\{T(k)-
{\int_k}\bigg[{{T({k_1})
T({k_2})}\over{2\alpha'{k_1}\cdot{k_2}+1}}
+ 2i{\mathcal{A}_\mu}({k_1})\times\nonumber\\
&\times&{{k_2^\mu}\over{\sqrt{\alpha'}{k_2^2}}}
\alpha({k_2})
+{{{\mathcal{A}_\mu}({k_1})
{\mathcal{A}_\nu}({k_2})}\over{2\alpha'{k_1}\cdot{k_2}-1}}\left({\eta^{\mu\nu}}-2\alpha'{k_2^\mu}{k_1^\nu}-{Q^{-2}}\alpha'
{k_1^\mu}{k_2^\nu}\right)+\nonumber\\
&+&{{{k_1}\cdot{k_2}}\over{\alpha'{k_1^2}{k_2^2}}}\alpha({k_1})
\alpha({k_2})\bigg]\bigg\},\nonumber\\
{\hat{\mathcal{A}}^\mu}\left[k,\varphi(\tau)\right]&=&\exp\left[-\alpha'{k^2}\varphi(\tau)\right]
\bigg\{{\mathcal{A}^\mu}(k)-2{\int_k}\bigg[{{T({k_1}){\mathcal{A}^\nu}({k_2})}\over{2\alpha'{k_1}\cdot{k_2}+1}}\left({\eta^\mu_\nu}+2\alpha'{k_2^\mu}{k_{1\nu}}\right)-\nonumber\\
&-&iT({k_1}){{k_2^\mu}\over{
\sqrt{\alpha'}{k_2^2}}}\alpha({k_2})\bigg]\bigg\},\nonumber\\
\hat{\alpha}\left[k,\varphi(\tau)\right]&=&{\mathcal{O}}(g),
\end{eqnarray}
where we have written ${\int_k}=-(g/2)\int{d^{26}}{k_1}\int{d^{26}}{k_2}{\delta^{26}}({k_1}+{k_2}-k)$ and also ${\mathcal{A}^\mu}(k)={A^\mu}(k)+
i{k^\mu}\alpha(k)/(\sqrt{\alpha'}{k^2})$.
Note that $\hat{\alpha}$ is a completely arbitrary ${\mathcal{O}}(g)$
function because we are free to add a
total boundary derivative to the 2D effective action.

At this stage it becomes clear that $T$, $A^\mu$ and $\alpha$ are not
the correct fields to define the $\beta$-functions because of the
mixing of $A^\mu$ and $\alpha$. Instead we must consider
$T$, $\mathcal{A}^\mu$ and $\alpha$. The correspondent anomalous dimensions 
are ${\lambda_{\hat{T}}}(k)=1-\alpha'{k^2}$,
${\lambda_{\hat{\mathcal{A}}}}(k)=-\alpha'{k^2}$ and
${\lambda_{\hat{\alpha}}}(k)=0$. Comparing with the solutions (\ref{6}) it is also clear that we
are missing the contribution associated with 
$\exp[({\lambda_k}+{\lambda_l})\varphi]$. This term is negligible when
compared with $\exp({\lambda_j}\varphi)$ if
${\lambda_k}+{\lambda_l}-{\lambda_j}<<0$. In the high energy region 
of phase space $-\alpha'{k_1^2}>>0$, $-\alpha'{k_2^2}>>0$,
$-2\alpha'{k_1}\cdot{k_2}>>0$ this is always true and since
$P({k_1},{k_2})+1\sim 2\alpha'{k_1}\cdot{k_2}$ the infrared
divergence is removed. Thus we define the $\beta$-functions by analytic
continuation from this region of momentum space. Our results are

\vspace{-0.75cm}
\begin{eqnarray}
{\beta_{\hat{T}}}(k)&=&(1-\alpha'{k^2})\hat{T}(k)+
{\int_k}\bigg[\hat{T}({k_1})\hat{T}({k_2})+{\hat{\mathcal{A}}_\mu}({k_1})
{\hat{\mathcal{A}}_\nu}({k_2})\left({\eta^{\mu\nu}}-2\alpha'{k_2^\mu}{k_1^\nu}\right)+\nonumber\\
&+&
2i{\hat{\mathcal{A}}_\mu}({k_1}){{k_2^\mu}\over{\sqrt{\alpha'}{k_2^2}}}\hat{\alpha}({k_2})\left(\alpha'{k^2}-1-\alpha'{k_1^2}\right)-{\alpha'\over{Q^2}}{k_1}\cdot\hat{\mathcal{A}}({k_1}){k_2}\cdot\hat{\mathcal{A}}({k_2})-\nonumber\\
&-&{{{k_1}\cdot{k_2}}\over{\alpha'{k_1^2}{k_2^2}}}\hat{\alpha}({k_1})\hat{\alpha}({k_2})\left(-\alpha'{k^2}+1\right)\bigg],\nonumber\\
{\beta_{\hat{\mathcal{A}}^\mu}}(k)&=&-\alpha'{k^2}{\hat{\mathcal{A}}^\mu}
(k)+2{\int_k}\bigg[\hat{T}({k_1}){\hat{\mathcal{A}}_\nu}({k_2})\left({\eta^{\mu\nu}}+2\alpha'{k_2^\mu}{k_1^\nu}\right)-\nonumber\\
&-&i\hat{T}({k_1}){{k_2^\mu}\over{\sqrt{\alpha'}{k_2^2}}}\hat{\alpha}({k_2})\bigg],\nonumber\\
{\beta_{\hat{\alpha}}}(k)&=&{\mathcal{O}}(g).            
\end{eqnarray}
Up to linear order the $\beta$-functions are invariant under the BRST gauge
transformation of the renormalized fields. With the introduction 
of the quadratic $\mathcal{O}(g)$ interactions this symmetry is dynamically broken.   

\section{The Effective Action and its Gauge
  Symmetry\newline Properties}
    
To order $g$ the $\beta$-functions define a
gradient flow in the theory space of couplings. The
associated effective action is equal to 

\vspace{-0.75cm}
\begin{eqnarray}
I&=&\int{d^{26}}X\bigg[\;{1\over{2}}\hat{T}(\alpha'\Box+1)
\hat{T}+{1\over{2}}{\hat{A}^\mu}\alpha'\Box{\hat{A}_\mu}-\sqrt{\alpha'}
{\hat{A}^\mu}{\partial_\mu}\hat{\alpha}-{1\over{2}}{\hat{\alpha}^2}-{g\over{3!}}{\hat{T}^3}-\nonumber\\
&-&{g\over{2}}\hat{T}{\hat{A}^\mu}{\hat{A}_\mu}-g\sqrt{\alpha'}\hat{T}{\hat{A}^\mu}
{\partial_\mu}\hat{\alpha}+{{g\sqrt{\alpha'}}\over{Q^2}}\hat{T}\hat{\alpha}\partial\cdot\hat{A}-g\alpha'\hat{T}{\partial_\nu}{\hat{A}^\mu}{\partial_\mu}
{\hat{A}^\nu}-\nonumber\\
&-&{{g\alpha'}\over{2{Q^2}}}\hat{T}\partial\cdot\hat{A}\partial\cdot\hat{A}-
{g\over{2{Q^2}}}\hat{T}{\hat{\alpha}^2}\bigg].\label{efa}
\end{eqnarray} 
This is only
possible if the total boundary derivative inducing an arbitrary
$\beta$-function for the ghost field $\hat{\alpha}$ is fixed to have the
necessary complementary terms to those shown in the $\beta$-functions
of $\hat{T}$ and $\hat{\mathcal{A}}$. Although at the linear order the
RG flow is associated with a unit metric in
the space of couplings, when the non-linear interaction terms are
introduced the metric develops an off-diagonal element
$G_{\hat{\mathcal{A}}\hat{T}}^\mu$,

\vspace{-0.25cm}
\begin{equation}
{G^{jl}}=\left[\begin{array}{ccc}
1& 0& 0\\
{G_{\hat{\mathcal{A}}\hat{T}}^\mu}& {\eta^{\mu\nu}}&0\\
0& 0& 1
\end{array}
\right],
\end{equation}
where

\vspace{-0.5cm}
\begin{equation}
{G_{\hat{\mathcal{A}}\hat{T}}^\mu}=-g\alpha'\left[{1\over{Q^2}}\partial\cdot
\hat{\mathcal{A}}\,{\partial^\mu}+\left(2+{1\over{Q^2}}\right){\partial^\mu}\partial\cdot\hat{\mathcal{A}}\right]{{(\alpha'\Box+1)}^{-1}}.
\end{equation}
This metric is invertible without the need to neglect
higher order terms. The inverse metric has the same form but with
${G_{\hat{\mathcal{A}}\hat{T}}^{\mu-1}}=-{G_{\hat{\mathcal{A}}\hat{T}}^\mu}$. 
With this condition on the invertibility of $G^{jl}$ the RG gradient
flow is unique.       

The BRST invariant kinetic terms in the effective action (\ref{efa}) are exactly those of WSFT. At the first non-linear
interaction level this symmetry is explicitly broken. To
see if any other invariance emerges we consider the variation of the interactions in Eq. (\ref{efa}) under the BRST transformation and study how new
quadratic gauge transformations might produce a compensating change of
the kinetic terms. We find that the effective action (\ref{efa}) is exactly invariant under the 
following $\gamma$-parameter family of non-linear gauge transformations

\vspace{-0.8cm}
\begin{eqnarray}
\delta\hat{T}&=& {g\over{2}}\left[\gamma\hat{\alpha}\Lambda+\sqrt{\alpha'}\left(2
{\hat{A}^\mu}{\partial_\mu}\Lambda-\gamma\Lambda\partial\cdot\hat{A}\right)\right],\nonumber\\
\delta{\hat{A}^\mu}&=&\sqrt{\alpha'}{\partial^\mu}\Lambda-{{g\sqrt{\alpha'}}\over{2}}\left[\gamma\Lambda{\partial^\mu}\hat{T}+(\gamma+2)\hat{T}{\partial^\mu}\Lambda\right],\nonumber\\
\delta\hat{\alpha}&=&\alpha'\Box\Lambda+{{g\gamma}\over{2}}\left(\hat{T}\Lambda-2\alpha'{\partial_\mu}\hat{T}{\partial^\mu}\Lambda-\alpha'\hat{T}\Box\Lambda\right).\label{11}
\end{eqnarray}

\vspace{-0.1cm}
\noindent Thought not equal in its detail to WSFT tree
level symmetry \cite{EW,KoS}, the $\sigma$-model invariance (\ref{11}) is of a similar
structural type. That is most notable in the transformation laws of $T$ and $A^\mu$ where the symmetry operator structure is the
same if $\gamma\not=0$. Missing in the $\sigma$-model are the exact operator
coefficients and smearing factors of Witten's
interactions. As is already clear in Eq. (\ref{efa}) it is when the ghost field
$\hat{\alpha}$ enters into the interactions that
the theories most differ.  This should be noted as evidence that when
probing deeper and deeper into the off-shell structure of this string
$\sigma$-model we will see it diverge more and more from WSFT. 

Thus it should not be too surprising that in the Feynman-Siegel gauge 
$\hat{\alpha}=0$ we are able to obtain Witten's structure of vertex
couplings \cite{EW,KoS} thought not its exact coefficients,

\vspace{-0.75cm}
\begin{eqnarray}
{I_{\mbox{\scriptsize FS}}}&=&\int{d^{26}}X\bigg[\;{1\over{2}}\hat{T}(\alpha'\Box+1)
\hat{T}+{1\over{2}}{\hat{A}^\mu}\alpha'\Box{\hat{A}_\mu}-{g\over{3!}}{\hat{T}^3}-{g\over{2}}\hat{T}{\hat{A}^\mu}{\hat{A}_\mu}-{{g\alpha'}\over{Q^2}}{\hat{A}_\nu}{\partial_\mu}
\hat{T}{\partial^\nu}{\hat{A}^\mu}-\nonumber\\
&-&g\alpha'\left(1+{1\over{2{Q^2}}}\right)\hat{T}{\partial_\nu}{\hat{A}^\mu}
{\partial_\mu}{\hat{A}^\nu}-{{g\alpha'}\over{2{Q^2}}}{\hat{A}^\mu}{\hat{A}^\nu}{\partial_\mu}{\partial_\nu}\hat{T}\bigg].
\end{eqnarray}

If we set $\partial\cdot A=\alpha$ the matter and ghost sectors of the
2D bare action decouple. Then the $\sigma$-model is
projected onto the theory space spanned by $T$ and $A^\mu$. The effective action is just Eq. (\ref{efa}) with 
$\partial\cdot\hat{A}=\hat{\alpha}$. The corresponding non-linear invariance 
follows from Eq. (\ref{11}) with $\gamma=0$, 

\vspace{-0.75cm}
\begin{eqnarray}
\delta\hat{T}&=&g\sqrt{\alpha'}{\hat{A}^\mu}{\partial_\mu}\Lambda,\nonumber\\
\delta{\hat{A}^\mu}&=&\sqrt{\alpha'}(1-g\hat{T}){\partial^\mu}\Lambda.
\end{eqnarray}
Here the
agreement with WSFT is almost
complete. Only missing in the interactions are its gaussian smearing
factors.

\section{Conclusions}

In this letter we have only considered the slice
of theory space corresponding to $T$, $A^\mu$ and $\alpha$. These are the relevant and marginal
perturbations about the on-shell gaussian fixed point. In a level
truncation type scheme \cite{KoS} they are the first levels in an
infinite tower of string states which in a fully off-shell description
must be considered together. The introduction of the higher massive fields can naturally be
done along the lines of this work but it is outside our present scope. We hope to report on progress in this direction in a future paper.

It is important to be aware of the limitations of the present
approach. Thought non-perturbative in $\alpha'$ the WFE does not allow us to probe very deeply
off-shell in the RG flow and is permissive to near mass-shell field
redefinitions \cite{AT,TB} which are bound to hide important physical
phenomena. In this
work we have introduced a purely off-shell ghost coupling and avoided the field redefinitions but
have not gone beyond the WFE. As a consequence our
results are just an indication (thought a strong one, we believe) 
that non-linear gauge symmetries and non-trivial string vacua of the 
type found in WSFT \cite{KoS} also exist in
the RG flow. Perhaps we might reach farther away with the Exact
Renormalization Group approach \cite{BM,HLP,EFE}. 

\section*{Acknowlegdements}

We would like to thank Yuri Kubyshin, Paul Mansfield and Tim Morris
for their support and encouragement. We also thank the Department of
Physics at Indiana University, U.S.A., for kind
hospitality in the early stages of this work.

\end{document}